\documentclass[preprint,superscriptaddress,preprintnumbers,showpacs]{revtex4}
\usepackage{amssymb}
\usepackage{graphicx}
\usepackage{dcolumn}
\usepackage{bm}
\usepackage{slashed}

\begin{document}

\newcommand*{\sjtu}{INPAC, SKLPPC and Department of Physics, Shanghai Jiao Tong University,  Shanghai, China}\affiliation{\sjtu}
\newcommand*{\shaoxin}{Department of Physics, Shaoxing University, Shaoxing,  Zhejiang, China }\affiliation{\shaoxin}
\newcommand*{\NTU}{CTS, CASTS and Department of Physics, \\National Taiwan University, Taipei, Taiwan}\affiliation{\NTU}
\newcommand*{\ncts}{Department of Physics, National Tsing Hua University, and National Center for Theoretical Sciences, Hsinchu, Taiwan}\affiliation{\ncts}

\hfill{}

\title{Lepton number violation and $h\to \gamma\gamma$ in\\ a radiative inverse seesaw dark matter model}
\affiliation{\NTU}
\author{Guan-Nan Li}\affiliation{\sjtu}
\author{Gang Guo}\affiliation{\sjtu}
\author{Bo Ren}\affiliation{\shaoxin}
\author{Ya-Juan Zheng}\affiliation{\NTU}
\author{Xiao-Gang He}\email{hexg@phys.ntu.edu.tw}\affiliation{\sjtu}\affiliation{\NTU}\affiliation{\ncts}

\begin{abstract}

We study phenomenological implications of a radiative inverse seesaw
dark matter model. In this model, because neutrino masses are
generated at two loop level with inverse seesaw, the new physics
mass scale can be as low as a few hundred GeV and the model also
naturally contain dark matter candidate. The Yukawa couplings
linking the SM leptons and new particles can be large. This can lead
to large lepton flavor violating effects. We find that future
experimental data on $\mu \to e \gamma$ and $\mu - e$ conversion can
further test the model. The new charged particles can  affect
significantly the $h \to \gamma \gamma$ branching ratio in the SM.
The model is able to explain the deviation between the SM prediction
and the LHC data. We also study some LHC signatures of the new
particles in the model.
\end{abstract}

\pacs{12.60.Fr,13.30.Ce, 95.35.+d, 14.80.Ec, 14.80.Fd}

\date{\today $\vphantom{\bigg|_{\bigg|}^|}$}

\maketitle

\section{Introduction}

Seesaw mechanism is one of the popular
mechanisms\cite{seesawI,seesawII,seesawIII} beyond the standard
model (SM) which can provide some explanations why neutrino masses
are small. Usually the seesaw scale is large making LHC study of the
new physics scale difficult. The inverse seesaw
mechanism\cite{inverse-seesaw} can lower the seesaw scale, because
in this type of models the  light neutrino masses are suppressed by
higher powers of new scale beyond the SM. If the inverse seesaw
mechanism is also achieved by radiative correction, the new scale
can be even lower. Such low new physics scale  can lead to large
testable effects in various experiments. Recently models of this
type have been proposed in which inverse seesaw mechanism is
radiatively realized at two loop level\cite{two-loop inverse
seesaw}. This allows the new physics scale to be in the hundreds GeV
range. To forbid tree and one loop level neutrino mass generation,
new unbroken symmetries are introduced. The lightest new particles
transforming non-trivially under the new symmetries are stable and
can play the role of  dark matter needed to explain about 23\% of
the energy budget of our universe\cite{dark-matter}.

In this paper we further study some phenomenologies in one of the
promising models. The model we will study is the $U(1)_D$ model
discussed in Ref.\cite{two-loop inverse seesaw}. There are several
new particles in this model. The large Yukawa couplings
linking the SM leptons and new particles in this model can have
large lepton flavor violating(LFV) effects. Future
experimental data on $\mu \to e \gamma$ and $\mu - e$ conversion can
further test the model. The new charged particles can  affect
significantly the $h \to \gamma \gamma$ branching ratio in the SM and the new contributions may be able to explain the
deviation between the SM prediction and the LHC data. We now provide
some details in the following.
\\

\section{ The model}

The model is based on the $SU(2)_L\times U(1)_Y$ SM electroweak gauge group with an unbroken global $U(1)_D$ symmetry.
The SM particles do not transform under the $U(1)_D$ symmetry. New particles in this model are vectorlike leptonic $SU(2)_L$
doublets $D_{L,R}$, two scalar singlets $S$, $\sigma$ and
a scalar $SU(2)_L$ triplet $\Delta$. Their SM and $U(1)_D$ charges are as follows
\begin{eqnarray}
D_{L,R}: (2,-1/2)(1)\;,\;\;\;\; S: (1,0)(-1)\;,\;\;\;\; \sigma: (1,
0)( 2)\;,\;\;\;\;\Delta: (3, -1)(2)\;.
\end{eqnarray}
In the above the two numbers in the first and the second brackets are the $SU(2)_L\times U(1)_Y$,
and the $U(1)_D$ quantum numbers, respectively.

The renormalizable terms for Yukawa couplings $L_D$  consistent with the symmetries of the model are
\begin{eqnarray}
L_D = - \bar L_L Y_D D_R S - \bar D_L M D_R - {1\over 2} \bar D_L Y_L D^c_L \Delta - {1\over 2} \bar D_R^c Y_R D_R \Delta^\dagger + h.c.
\end{eqnarray}

The allowed renormalizable terms in the potential $V_D$ are given by
\begin{eqnarray}
V_D &=& -\mu^2_H H^\dagger H + \lambda_H (H^\dagger H)^2 + \mu^2_S S^\dagger S + \lambda_S (S^\dagger S)^2 + \mu^2_\sigma \sigma^\dagger \sigma + \lambda_\sigma (\sigma^\dagger \sigma)^2\nonumber\\
& + & \mu^2_\Delta \Delta^\dagger \Delta + \lambda^\alpha_\Delta (\Delta^\dagger \Delta \Delta^\dagger \Delta)_\alpha + \sum_{ij}\lambda_{ij} i^\dagger ij^\dagger j
+ (\mu_{S\sigma} S^2 \sigma  + \lambda_{\Delta \sigma H} H \Delta \sigma^\dagger H + h.c.),
\end{eqnarray}
where the sum $\sum_{ij}$ is over all possible $i$ and $j$, and $i$
to be one of the $H$, $S$, $\sigma$ and $\Delta$. The allowed terms
are
\begin{eqnarray}
&&\lambda^\beta_{H\Delta} (H^\dag H \Delta^\dag \Delta)_\beta + \lambda_{H\sigma} (H^\dag H \sigma^\dag \sigma)+ \lambda_{HS} (H^\dag H S^\dag S)\nonumber\\
&&+ \lambda_{\Delta S} (\Delta^\dag \Delta S^\dag S) +
\lambda_{\Delta \sigma} (\Delta^\dag \Delta \sigma^\dag \sigma) + \lambda_{\sigma S} (\sigma^\dag \sigma S^\dag S)\;.
\end{eqnarray}
In the above the indices $\alpha$ and $\beta$ indicate different ways of forming singlets. They are given by
\begin{eqnarray}
  (\Delta^\dag \Delta \Delta^\dag \Delta)_1 = \Delta^*_{ij} \Delta_{ij} \Delta^*_{kl} \Delta_{kl}\;,\;\;\;\;(\Delta^\dag \Delta \Delta^\dag \Delta)_2 = \Delta^*_{ij} \Delta_{ik} \Delta^*_{kl} \Delta_{jl} \cr
  (\Delta^\dag \Delta H^\dag H)_1 = \Delta^*_{ij} \Delta_{ij} H^*_{k} H_{k}\;,\;\;\;\;(\Delta^\dag \Delta H^\dag H)_2 =  \Delta^*_{ij} \Delta_{kj} H^*_{k} H_{i}
\end{eqnarray}

If both $S$ and $\Delta$ develop non-zero vev's, the Lagrangian
$L_D$ will give the usual inverse seesaw masses to neutrinos. In
that case there will be a Goldstone boson due to breaking of the
global $U(1)_D$ symmetry which may be problematic. To avoid the
appearance of massless Goldstone boson in the theory, a possible
approach is to keep the global symmetry to be exact and therefore no
Goldstone boson emerges. This requires $\mu^2_i$ to be  all larger than
zero. This also forbids the light neutrinos to have non-zero masses
at tree level. However, Majorana neutrino masses can be generated at
two loop level through the Feynman diagram shown in Fig.1.

\begin{figure}
\includegraphics[width=6cm]{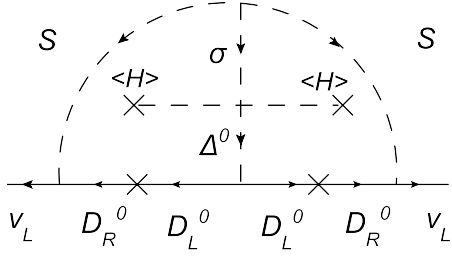}
\caption{Two loop Feynman diagram for neutrino mass generation.}\label{two-loop}
\end{figure}

Carrying out the loop integrals, one obtains neutrino mass matrix
$m_\nu$ in the bases where $M$ is diagonal
\begin{eqnarray}
  m_\nu^{ij} =\frac{v_H Y_D^{ik}(\lambda_{\Delta\sigma H}\mu_{S\sigma} Y_L^{kl}) Y_D^{jl}  v_H}{M^2_{kk}}\kappa_{kl}\;, \label{mass}
  \end{eqnarray}
 where $\kappa_{kl}$ is defined as
 \begin{eqnarray}
 \kappa_{kl} &=&\delta_{kl}\frac{1}{2(4\pi)^4}{1\over (1- m^2_S/M^2_{kk})^2}[g(m_{\phi_1},m_S,m_S)-g(m_{\phi_1},M_{kk},m_S) \cr
 &&-g(m_{\phi_1},m_S,M_{kk})+g(m_{\phi_1},M_{kk},M_{kk})]\;.\label{nu-mass}
 \end{eqnarray}
 \begin{eqnarray}
  g(m_1,m_2,m_3)=\int^{1}_{0} dx [1+Sp(1-\mu^2)-\frac{\mu^2}{1-\mu^2}\log\mu^2]\nonumber
  \end{eqnarray}
  with $\mu^2=\frac{ax+b(1-x)}{x(1-x)}, a=\frac{m^2_2}{m^2_1}, b=\frac{m^2_3}{m^2_1}$.
$Sp(z)$ is the Spence function or the dilogarithm function
  \begin{eqnarray}
  Sp(z)=-\int^z_0 {\ln(1-t)\over t} dt
  \end{eqnarray}
In the above we have assumed that $\sigma$ and the neutral component of $\Delta$ have almost equal mass $m_{\phi_1}$.

There are candidates for dark matter in this model. The neutral
heavy particles in $D_{L,R}$ and $\Delta$ have non-zero hypercharges
and have problems to play the role of dark matter. The natural dark
matter candidate field is $S$. It does not have a non-zero
hypercharge and does not mix with any particles with hypercharge
($\sigma$ mixes with $\Delta$). As long as dark matter properties
are concerned, this model is very similar to the real singlet
(darkon) model\cite{real-singlet} and therefore has similar dark
matter properties\cite{darkon, tc} and is identical to the complex
scalar singlet model\cite{complex-singlet} with degenerate mass for
the real and imaginary parts of S. The term $S^\dagger S H^\dagger
H$ is important for dark matter relic density and direct detection
studies.

The Higgs boson $h$ properties, its mass and its couplings to SM
particles (fermions and gauge bosons), are the same as those in  the
SM at the tree level. The recent LHC data indicate that the mass is
about 126 GeV\cite{LHC-higgs} which can be applied to this model. It
has been shown that the dark matter relic density and direct
detection constraints can be simultaneously satisfied with
appropriate dark matter mass. The range of a few tens of GeV for
dark matter mass is in trouble. However, dark matter mass about half
of the Higgs mass or larger than 130 GeV is allowed\cite{two-loop
inverse seesaw}. In our later discussions, we will take $m_S = 150$
GeV  for illustration.

\section{Neutrino masses and LFV}

The formula in Eq.(\ref{nu-mass}) determines whether the model is
consistent with current data on neutrino mixing and
masses\cite{neutrino-data}. In order to have at least two neutrinos
with non-zero mass to be consistent with data, more than one generation of $D_{L,R}$ are
needed. We will assume that there are three of them. The mixing
pattern is determined by two Yukawa couplings, $Y_D$ and $Y_L$. With
three $D_{L,R}$, they both are $3\times 3$ matrices. In our
numerical calculations, we will assume that the flavor structure is
determined by the Yukawa coupling $Y_D$ with $Y_D = y_D U_{PMNS}\hat
Y_D$ with $Y_L$ diagonal for both normal and inverted hierarchies
for neutrino masses. In our later calculations we will use the
central values from recent global fit data in
Ref.\cite{neutrino-data} for neutrino mixing angles and mass squared
differences for both normal and inverted hierarchies (NH and IH) for
our discussions
\begin{eqnarray}\label{fitdata}
&&\sin^2\theta_{12}=0.307^{+0.018}_{-0.016} (\rm NH,IH);
\sin^2\theta_{23}=0.386^{+0.024}_{-0.021}({\rm NH}),0.392^{+0.039}_{-0.022}({\rm IH});\nonumber\\
&&\sin^2\theta_{13}=0.0241\pm0.0025({\rm NH}),0.0244^{+0.0023}_{-0.0025}({\rm IH});\nonumber\\
&&\delta m^2=m^2_2 - m^2_1=(7.54^{+0.26}_{-0.22})\times 10^{-5}{\rm eV}^2
({\rm NH,IH});\nonumber\\
\nonumber &&|\Delta m^2| = |m^2_3-(m^2_2+m^2_1)/2|=(2.43^{+0.06}_{-0.1})\times
10^{-3}{\rm eV}^2 ({\rm NH}),(2.42^{+0.07}_{-0.11})\times 10^{-3}{\rm eV}^2 ({\rm IH});\\
&&\delta = 194.4^\circ  ({\rm NH}),~~196.2^\circ ({\rm IH}).
\end{eqnarray}

In the following we show two sets of model parameters which can fit
known data for neutrinos and take them as bench mark values.

For the normal hierarchy, choosing $\hat Y_D = diag(1, \sqrt{1.03},
\sqrt{1.77})$, $y_D \times \lambda_{\Delta\sigma H} =10^{-3}$, $Y_L
= \mbox{I} \times 10^{-2}$, $\mu_{S\sigma} = 100$GeV, $m_{\phi_1} =
300$GeV, $m_S = 150$GeV, $M_{ii}=500$GeV, we can get all the three
neutrino mass $3.39\times 10^{-2}$eV, $3.50\times 10^{-2}$eV,
$5.98\times 10^{-2}$eV, respectively. These are consistent with
data.

For inverted hierarchy case, we just need to replace $\hat Y_D$ with
$\hat Y_D = diag(\sqrt{1.46}, \sqrt{1.48}, \sqrt{0.100})$, with all
the other parameters unchanged, the neutrino masses will be
$4.93\times 10^{-2}$eV, $5.01\times 10^{-2}$eV, $3.39\times
10^{-3}$eV, respectively. Again, these numbers are consistent with
data.

For neutrino masses, the two parameters $y_D$ and $\lambda_{\Delta
\sigma H}$ appear together, but for charged lepton LFV processes
which happen at one loop level, they only depend on $y_D$. We will
study how $y_D$ is constrained by data from $l_i\to l_j\gamma$ and
$\mu - e$ conversion.

\begin{figure}[!h]
  \centering
 \includegraphics[width=12cm]{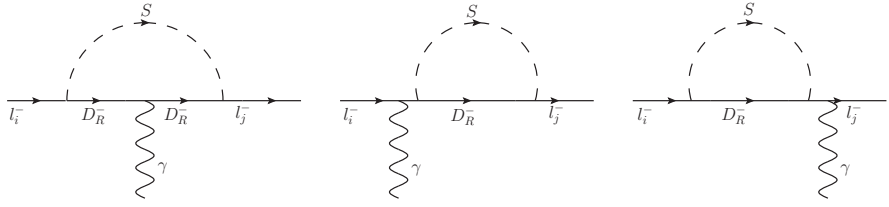}
    \caption{Feynman diagrams for $l_i \to l_j\gamma$. }
    \label{a}
\end{figure}
\begin{figure}[!h]
    \centering
    \includegraphics[width=12cm]{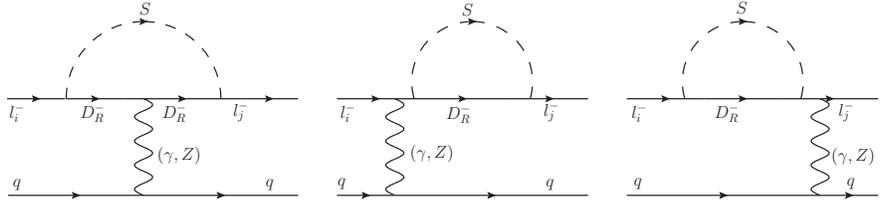}
    \caption{Feynman diagrams for $l_i - l_j$
conversion. }
    \label{b}
\end{figure}

Radiative leptonic decay $l_i \to l_j\gamma$ can occur at one loop
level as shown in Fig.\ref{a}. By attaching $\gamma$, and changing
$\gamma$ into $Z$, and then let $\gamma$ and $Z$ connect to quark, as
shown in Fig.\ref{b},  $\mu - e$ conversion can be induced.  For our
case the Lagrangian responsible for $l_i\to l_j\gamma$ and $l_i -
l_j$ conversion can be written as
\begin{eqnarray}
 {\cal L} = &-& \bar l_j \sigma^{\mu\nu} (A_{Lji} P_L + A_{Rji} P_R) l_i F_{\mu\nu} + [\sum_q eQ_q\bar q \gamma^\mu q \bar l_j B_{Lji}\gamma_\mu P_L l_i + H.c.]\;,
\end{eqnarray}
and the functions $A_{L,R}$ and $B_L$ are given by
\begin{eqnarray}
&&A_{Lji}=Y_{Djk} Y^*_{Dki} \frac{e}{32 \pi^2}\frac{1}{m_S^2}
F_D(\frac{M_k^2}{m^2_S}) m_j\;,\;\;A_{Rji} = {m_i\over
m_j}A_{Lji},\nonumber
\\
&&B_{Lji}=Y_{Djk}Y^*_{Dki}{e \over 16 \pi^2}{1\over m_S^2}G_D({M_k^2 \over m^2_S})\;,\nonumber\\
&&F_D(z)=\frac{z^2-5z-2}{12(z-1)^3}+\frac{z \ln z}{2(z-1)^4}\;,\nonumber\\
&&G_D(z)= {7 z^3-36z^2+45z-16+6(3z-2)\ln z \over 36(1-z)^4}\;.
\label{fcnc}
\end{eqnarray}

The LFV $l_i \to l_j \gamma$ decay branching ratio is easily evaluated
by
\begin{eqnarray}
B(l_i \to l_j \gamma) = \frac{48\pi^2}{G_F^2 m^2_i} (|A_{Lji}|^2 +
|A_{Rji}|^2).
\end{eqnarray}
The strength of $\mu - e$ conversion is measured by the quantity,
$B^A_{\mu \to e}= \Gamma^A_{conv}/\Gamma^A_{capt} = \Gamma(\mu^- +
A(N,Z) \to e^- +A(N,Z))/\Gamma(\mu^- + A(N,Z) \to \nu_\mu + A(N+1,
Z-1))$.  To obtain the conversion rate, one needs to convert the
quarks in Eq.(\ref{fcnc}) into relevant nuclei. We will use the
theoretical values compiled in Ref.\cite{Ref:Mu2Ecv1}. We
have\cite{Ref:meco4g}
\begin{eqnarray}
{B^A_{\mu\to e}\over B(\mu \to e\gamma)} = R^0_{\mu\to e}(A) \left |
1 + {\tilde g^{(p)}_{LV} V^{(p)}(A)\over A_R D(A)} + {\tilde
g^{(n)}_{LV} V^{(n)}(A)\over A_R D(A)}\right |^2\;,
\end{eqnarray}
where
\begin{eqnarray}
R^0_{\mu\to e}(A) = {G^2_F m^5_\mu \over 192 \pi^2
\Gamma^A_{capt}}|D(A)|^2\;.
\end{eqnarray}
and
\begin{eqnarray}  \tilde g^{(p)}_{LV} = 2 g_{LV(u)}+g_{LV(d)},
\tilde g^{(n)}_{LV} = g_{LV(u)}+ 2 g_{LV(d)}, g_{LV(q)}=-4eQ_q
m_{\mu} B_L.
\end{eqnarray}
The parameters $D(A)$, $V^{(p,n)}(A)$ are nuclei dependent
quantities. Several of them are given in Ref.\cite{Ref:Mu2Ecv1}.

\begin{figure}[!h]
\centering
\includegraphics[width=7cm]{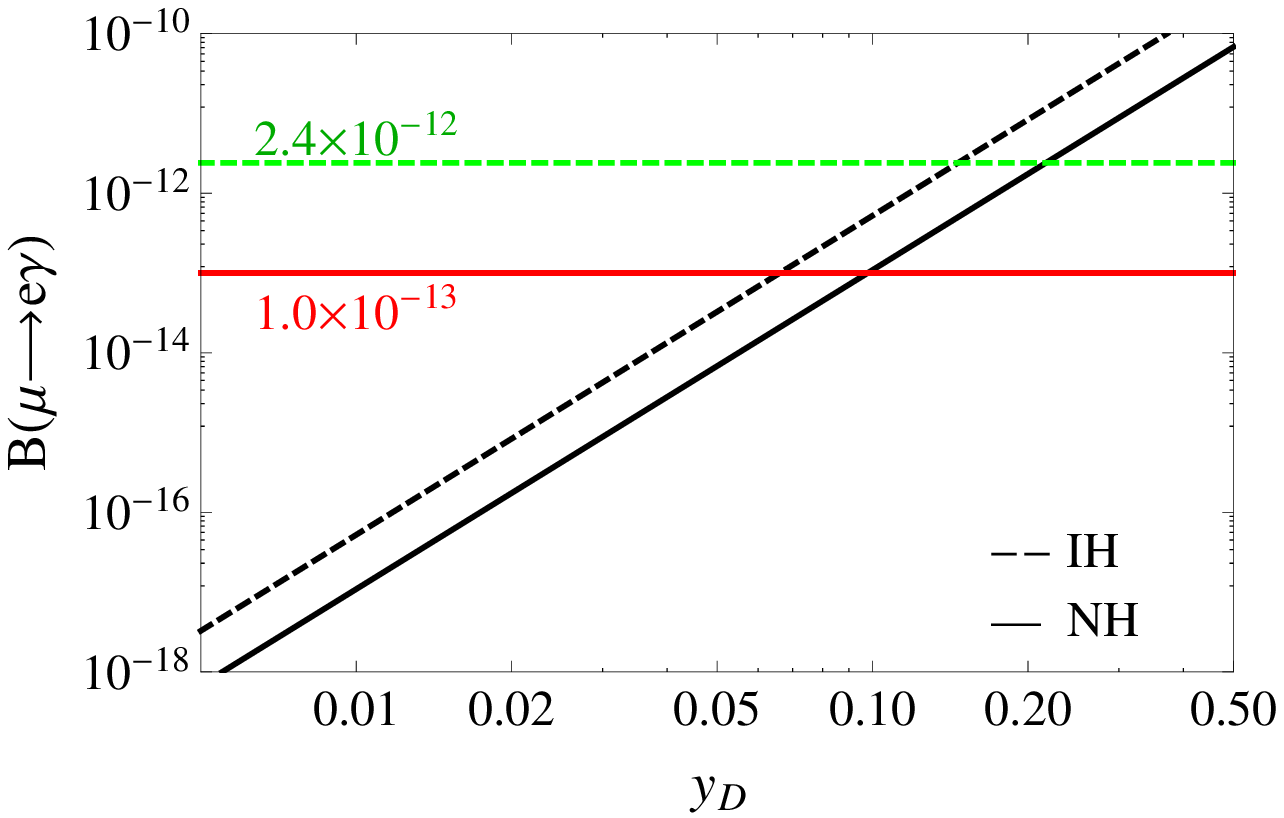}
\hspace*{2em}
\includegraphics[width=7cm]{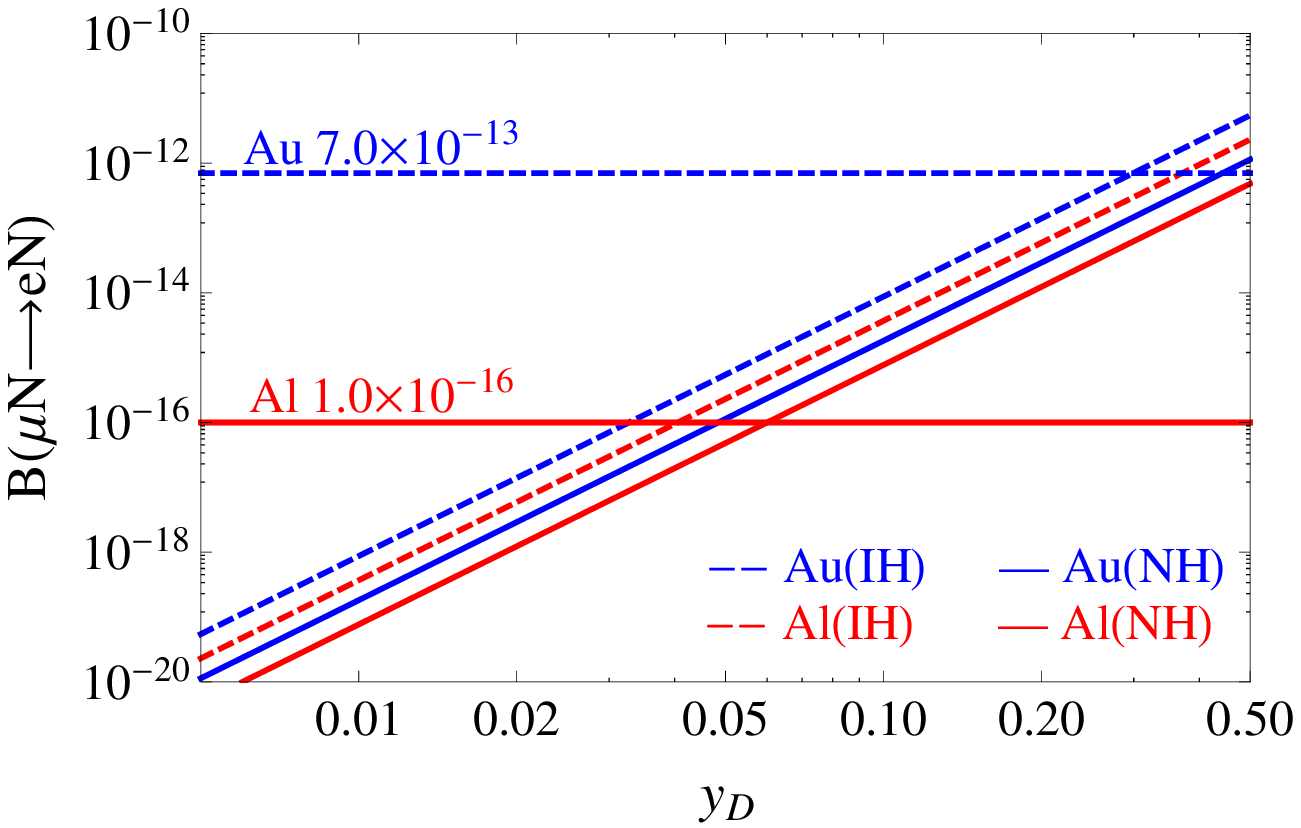}
\includegraphics[width=7cm]{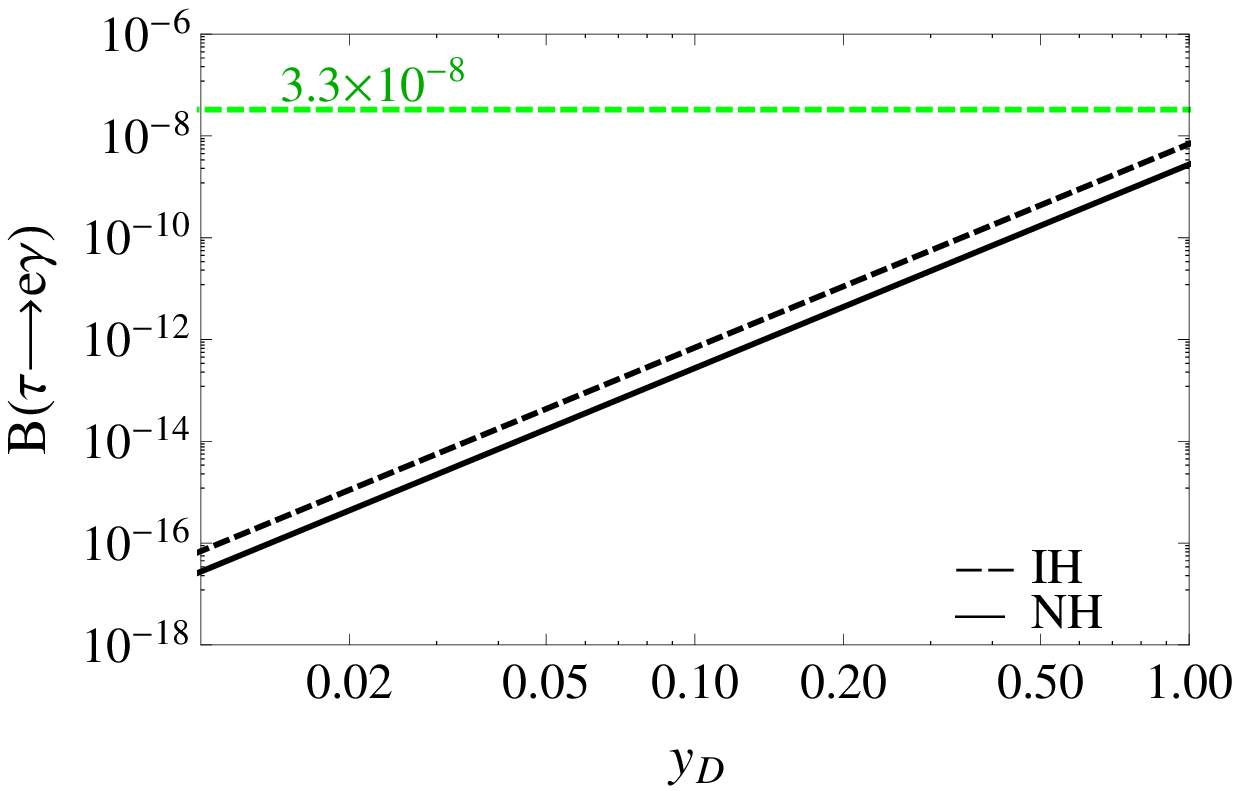}
\hspace*{2em}
\includegraphics[width=7cm]{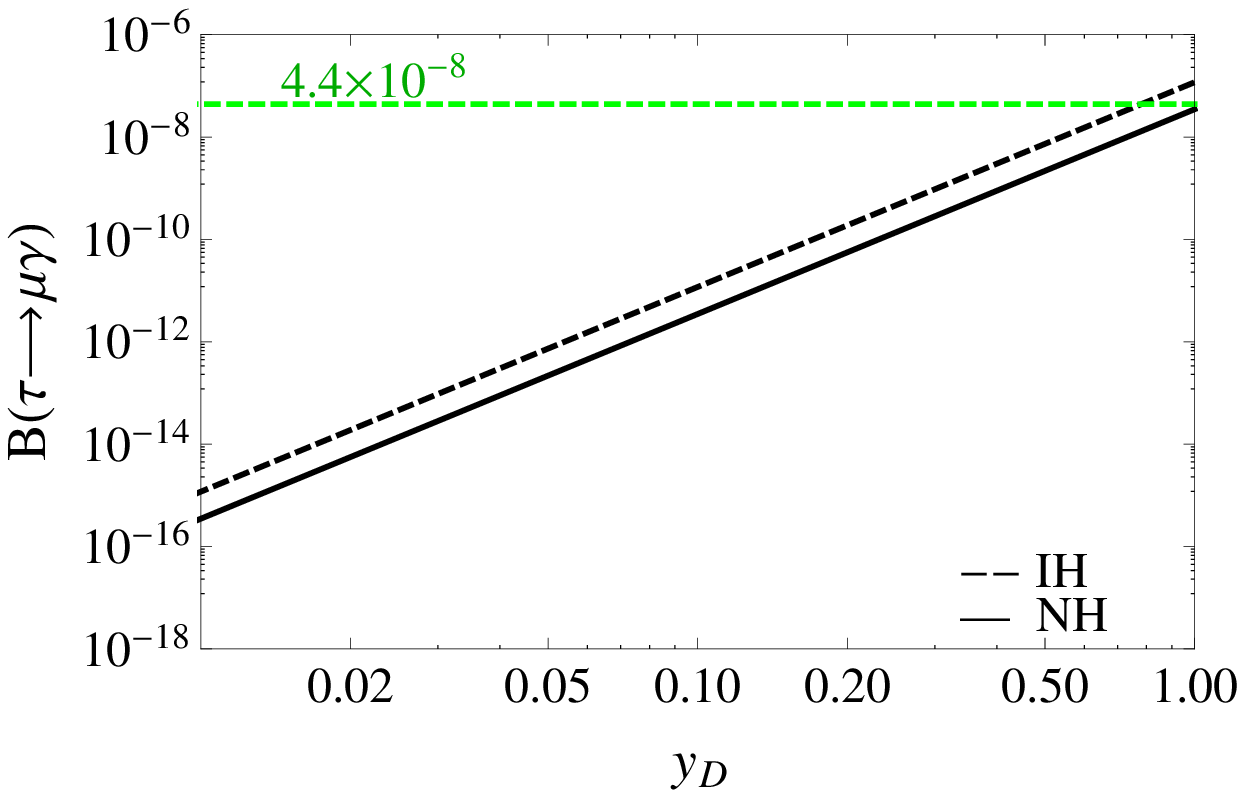}
\caption{Figures on the top show constraints on $y_D$ from
$B(\mu  \to e \gamma)$ and $\mu - e$ conversion rate
with current bound (dashed line) and future sensitivity (solid line)
for normal(solid line) and inverted(dashed
line) hierarchy. Figures at the bottom show constraints on $y_D$ from $\tau \to e\gamma$ (left) and
$\tau\to \mu \gamma$ (right).  } \label{yd}
\end{figure}

With the bench mark values for the model parameters fixed, the
$B(l_i \to l_j\gamma)$ and $\mu - e$ conversion rate are all dependent
on the coupling constant $y_D$. We now discuss the constraint on
$y_D$.

Although $\mu \to e\gamma$ has not been observed, there are
stringent constraint on the upper limit of the branching ratio. The
current upper limit is $2.4\times 10^{-12}$  at the 90\%
c.l.\cite{meg-new}. Experimental sensitivity will be improved. We
take $B(\mu \to e\gamma)=1\times 10^{-13}$\cite{Ref:MEG-pot} as the
near future improved MEG experimental sensitivity to constrain the
parameter $y_D$. There are also bounds for the process of $\tau \to
\mu(e)\gamma$. The current 95\% c.l. experiment bounds are $B(\tau
\to e\gamma) < 3.3 \times10^{-8}$, $B(\tau \to \mu\gamma) < 4.4
\times10^{-8}$\cite{tau to mu e gamma}.

There are several measurements of $\mu - e$ conversion on various
nuclei. The best experimental bound is from Au nuclei
with the 90\% c.l. upper bound given by $B^{\text{Au}}_{\mu \to e} < 7\times
10^{-13}$\cite{Ref:Mu2E-Au}. For Au, the relevant parameters
determined by method I in Ref.\cite{Ref:Mu2Ecv1} are given by:
$D(\text{Au})=0.189$, $V^{(p)}(\text{Au})=0.0974,
V^{(n)}(\text{Au})=0.146$ and $R^0_{\mu\to
e}(\text{Au})=0.0036$\cite{Ref:Mu2Ecv1}. There are several planned experiments, such as Mu2E\cite{Ref:Mu2E}/COMET\cite{Ref:COMET}
for $\mu -e $ conversion using Al. The sensitivities are expected to
reach $10^{-16}$\cite{Ref:COMET}. For Al nuclei, the relevant
parameters for our calculations are given by  $D(\text{Al})=0.0362$,
$V^{(p)}(\text{Al})=0.0161, V^{(n)}(\text{Al})=0.0173$ and
$R^0_{\mu\to e}(\text{Al})=0.0026$\cite{Ref:Mu2Ecv1}.

The constraints on $y_D$ are shown in Fig.\ref{yd}. We see that the
current upper limits from $\mu \to e \gamma$ and $\mu - e$
conversion using Au can already constrain $y_D$ to be less than 0.2
and 0.4, respectively. Future $\mu -e$ conversion experiments can
reach  a sensitivity of 0.05 on $y_D$. The model will be constrained
when new data become available. The constraints from $\tau \to \mu
(e) \gamma$ are weaker.

 In the above studies, we have taken some bench mark values to have
some ideas about the possibility of observing LFV effects. Our
studies show that it is possible to have large observable LFV
effects  at near future experiment, in particular for $\mu - e$
conversion experiments. We, however, should note that from such
studies it is not possible to rule out the model because the allowed
parameters can have large or small LFV effect. For example,  for
neutrino mass generation, the scale of neutrino mass depends on $y_D
\times \lambda_{\Delta\sigma H}$, but  $\lambda_{\Delta\sigma H}$
does not show up directly in the leading LFV effects which we have
studied. Even assuming all other relevant parameters are fixed in
the model, by adjusting the size of $\lambda_{\Delta\sigma H}$ one
can have different values of $y_D$ to satisfy constraint on $y_D$
from LFV processes. There are some other processes which can provide
additional tests for the model. We find that the correlation of $h
\to \gamma\gamma$ and $h \to \gamma Z$ can be a good indicator. We
will discuss this later.

In the above studies, the new physics scale is set by the masses
of particles in the $D$ doublet. We now briefly discuss LHC
signature of these particles. $D^0 \bar D^0$, $D^0 D^-$ and $D^-
D^+$ can be pair produced through $Z$, $W^-$, $\gamma$ and $Z$
s-channel exchanges with the cross sections of order ${\cal O} (10)$
fb for mass $m_D$ around 500 GeV.

For $D^0 \bar D^0$ production, since $D^0$ decays into $\nu S$, the
signature is missing energy which would be similar to dark matter
signature with photon or gluon emissions from the initial quark. The
final state is thus a high-$p_T$ photon/gluon and missing energy.
Detections of dark matter pair production processes from
CMS\cite{CMS_DM} and ATLAS\cite{ATLAS_DM} have been performed. As no
excess from SM predictions observed in both experiments, constraints
on dark matter mass can then be given accordingly for pair
production of dark matter candidates\cite{DM_constraints}. The
current data cannot rule out $D^0$ of order a few hundred GeV.

$D^-$ will decay into $l^- S$. The pair production of $D^0 D^-$
through $W^-$ exchange can be searched by $pp \to l^-+\slashed{E_T}
+ X$. There will be SM background from $W^-\to l^-\bar{\nu}$ and
$W^-Z/W^-\to l^- \bar{\nu} \nu\bar{\nu}$. Additionally, $W^+W^-$
production with leptonic decays will also have a possibility to be background
 when one charged lepton (here $l^+$) is too soft or too forward. To optimize
the signal from these backgrounds, one can impose $p_T^l$ cut on
charged lepton.
With 8 TeV energy at the LHC and $20 fb^{-1}$, it is difficult to cut down the background
to have enough signal events. We find that it is possible to achieve a discovery level at 5$\sigma$ for 14 TeV center of mass frame enery with $300 fb^{-1}$.
In Table \ref{tab:cs}, we show the cross section
with a selective cut of $p_T^l > 120$ GeV. We have chosen the cut
for $p_T^l$ so that the signal can be established at 5$\sigma$ level
statistically. With a higher cut for $p_T^l$, one can have a higher
significance level, but the event number will be smaller. With this
cut of $p_T^l$  the signal is slightly eliminated while the
backgrounds are effectively suppressed. Note that, in the
calculation of signal, we have taken the  $D$ doublets with almost
degenerate masses and used the bench mark values given before. In
the narrow width approximation, the cross section for $l$ charged
lepton in the final state is proportional to $\sum_i
(Y_D^{li}Y^{li*}_D/\sum_l Y_D^{li}Y^{li*}_D)$ for degenerate $D_i$.
The cross section for $pp \to l^-+\slashed{E_T} + X$, therefore, is
almost independent of $y_D$. The pair production of $D^-D^+$ will
have the signal $l^-l^+$ plus missing energies. Since there are two
leptons in the final state, the analysis is more involved. We will
not discuss this here. The numbers in Table \ref{tab:cs} show that
with an integrated luminosity of $300 {\rm fb}^{-1}$ for
centre-of-mass energy at 14 TeV,  for both normal and inverted
hierarchy cases, signal of $5\sigma$ significance can be achieved
using $pp \to l^-+\slashed{E_T} + X$ .  It is interesting to carry
out such a search.

\begin{table}[h!]
\begin{centering}
\begin{tabular}{|c|c|c||c|c|c|}
\hline
&\multicolumn{2}{c||}{Signal [fb](NH and IH)} & \multicolumn{3}{c|}{Background[fb] }\tabularnewline
\hline
&\multicolumn{2}{c||}{$\sigma(pp\to D^0Sl^{-})$}& $\sigma(W^-\to l^-\bar{\nu})$&$\sigma(W^-Z/W^-\to l^-\bar{\nu}\nu\bar{\nu})$ &$\sigma(W^+W^-)$ \tabularnewline
\hline
14~{\rm TeV}&~~~~11.1~~~~~&14.2&$8.67\times 10^6$&345.3&1856\tabularnewline
\hline
$p_T^l >$120 {\rm GeV} &9.66&12.3& 1080&6.78&26\tabularnewline
\hline
\end{tabular}\caption{Cross sections of signal with $m_D=500{\rm GeV}$ and $m_S=150~{\rm GeV}$ and corresponding backgrounds. In both signal and backgrounds, charged lepton of $e^-$ and $\mu^-$ are included.}
\label{tab:cs}
\par\end{centering}\end{table}

\section{$h \to \gamma\gamma$}

There are strong indications from LHC that the Higgs particle has
been discovered with a mass of 126 GeV whose couplings to gauge
bosons are consistent with SM Higgs, but with an enhanced $h\to
\gamma\gamma$ branching ratio. The experimental
value\cite{LHC-higgs} for this channel is $1.8\pm 0.5$ (ATLAS)
($1.56\pm 0.43$(CMS)) times that predicted by the SM. Recently ATLAS
has updated their result with\cite{atlas-1} $1.8 \pm0.3 ({\rm
stat.}) ^{+0.21}_{-0.15}({\rm sys.}) ^{+0.20} _{-0.14} ({\rm
theory})$ times the value predicted by the SM. The
central value is higher than the SM prediction. If confirmed, new
physics beyond the SM is required to explain it. In the model we are
studying, this can be explained by new contribution from charged particles in the
triplet scalar $\Delta$ with relatively low mass coupled to the
usual Higgs boson at loop levels. We now discuss how enhancement can
be achieved. There have been extensive studies for similar triplet
scalar contributions to $h \to \gamma\gamma$\cite{triplet scalar for
h to rr, geng}.  Our study is more model inspired, the triplet does not
have non-zero vev, and also the $\Delta$ does not decay into pure SM
particles. The LHC signatures for $\Delta$ particles are different
than other models.

In the model we are considering, electroweak symmetry breaking is induced by the non-zero vev of Higgs doublet $H = (h^+, (v+h + i I)/\sqrt{2})^T$ . The charged $h^+$ and the neutral fields $I$ are ``eaten'' by $W$ and $Z$. The $h$ is the physical Higgs field similar to the one in SM. Since this is the only field having a non-zero vev in the theory, at the tree level, the Higgs $h$ couplings to gauge bosons are the same as those in the SM. The Yukawa couplings to SM fermions also have the same form as those in the SM. At one loop level, deviations start to show up. A particularly interesting one is modification for $h \to \gamma\gamma$ coupling, due to the existence of new charged particles $\Delta^{-,--}$ and their non-zero couplings to $h$. Note that the new particles, do not have strong interactions, the process $gg\to h$ is not affected to the lowest order. So the model  will not alter the production rate of $h$ predicted by the SM to the leading order in agreement with data.

The couplings of $h$ to $\Delta^{-,--}$ come from  $\lambda^{1,2}_{\Delta H}(\Delta^\dag \Delta H^\dag H)_{1,2}$ after $H$ develops vev. The $h \bar \Delta \Delta$ couplings are given by
\begin{eqnarray}
L \sim -[\lambda^1_{\Delta H} (\Delta^+ \Delta^- + \Delta^{++}
\Delta^{--})  + \lambda^2_{\Delta H}  ( \Delta^{++} \Delta^{--} +
{1\over 2}\Delta^+ \Delta^-)]v h\;.
\end{eqnarray}

Combined with contributions from $W$ and top in the loop, the $h\to \gamma \gamma$ rate is modified by
a factor $R_{\gamma\gamma} = \Gamma(h\to \gamma \gamma)_{U(1)_D}/\Gamma(h \to \gamma\gamma)_{\rm SM}$ given by
 \begin{eqnarray}
 &&R_{\gamma\gamma} = \vert 1+ {v^2 \over 2} {1 \over A_1(\tau_W) + N_c Q_t^2 A_{1/2}(\tau_t)} \{{\lambda^1_{H\Delta} + {1\over 2} \lambda^2_{H\Delta}\over  m_{\Delta^-}^2} A_0(\tau_{\Delta^-}) + {4(\lambda^1_{H\Delta} + \lambda^2_{H\Delta}) \over  m_{\Delta^{--}}^2} A_0(\tau_{\Delta^{--}}) \}    \vert^2 \nonumber\\
 \label{RR}
 \end{eqnarray}
 where $\tau_i \equiv (m^2_h/4m_i^2 ),~i=t,W,\Delta^- ~{\rm and}~ \Delta^{--}$. $N_c$ is the degree freedom of color and $Q_t$ is the charge of top quark. $A_1(\tau_W)$ and $A_{1/2}(\tau_t)$ come from SM $W$ boson and top quark contributions. $A_0(\tau_\Delta)$ comes from
 new scalars in the model. They are given by
 \begin{eqnarray}
&&A_0(x) = -x^{-2} [x-f(x)];
A_{1/2}(x)=2x^{-2}[x+(x-1)f(x)];\nonumber\\
&&A_1(x) =-x^{-2}[2x^2+3x+3(2x-1)f(x)];\nonumber\\
&&f(x)=\left\{ \begin{array}{l}
          \arcsin^2 \sqrt{x},  ~~ x \ge 1 \\
          -{1\over 4} [\ln { 1+\sqrt{1-x^{-1}} \over 1-\sqrt{1-x^{-1}}}-i \pi]^2, ~~x < 1
          \end{array}\right.
\end{eqnarray}

Eq.(\ref{RR}) tells that new contributions to the ratio
$R_{\gamma\gamma}$  depend on not only the couplings
$\lambda_{H\Delta}^{1,2}$, but also the masses of the charged
scalars.  The scalar masses depend on several parameters. Neglecting
the mixing between $\sigma$ and $\Delta^0$, the component fields in
$\Delta$ masses are given by
\begin{eqnarray}
&&m^2_{\Delta^0} = \mu^2_\Delta + {1\over 2} \lambda^1_{H\Delta} v^2\;,\nonumber\\
&&m^2_{\Delta^-} = \mu^2_\Delta + {1\over 2} \lambda^1_{H\Delta} v^2 + {1\over 4} \lambda^2_{H\Delta} v^2\;,\\
&&m^2_{\Delta^{--}} = \mu^2_\Delta + {1\over 2} \lambda^1_{H\Delta} v^2 + {1\over 2} \lambda^2_{H\Delta} v^2\;,\nonumber
\end{eqnarray}

\begin{figure}[!h]
\centering
\includegraphics[width=7cm]{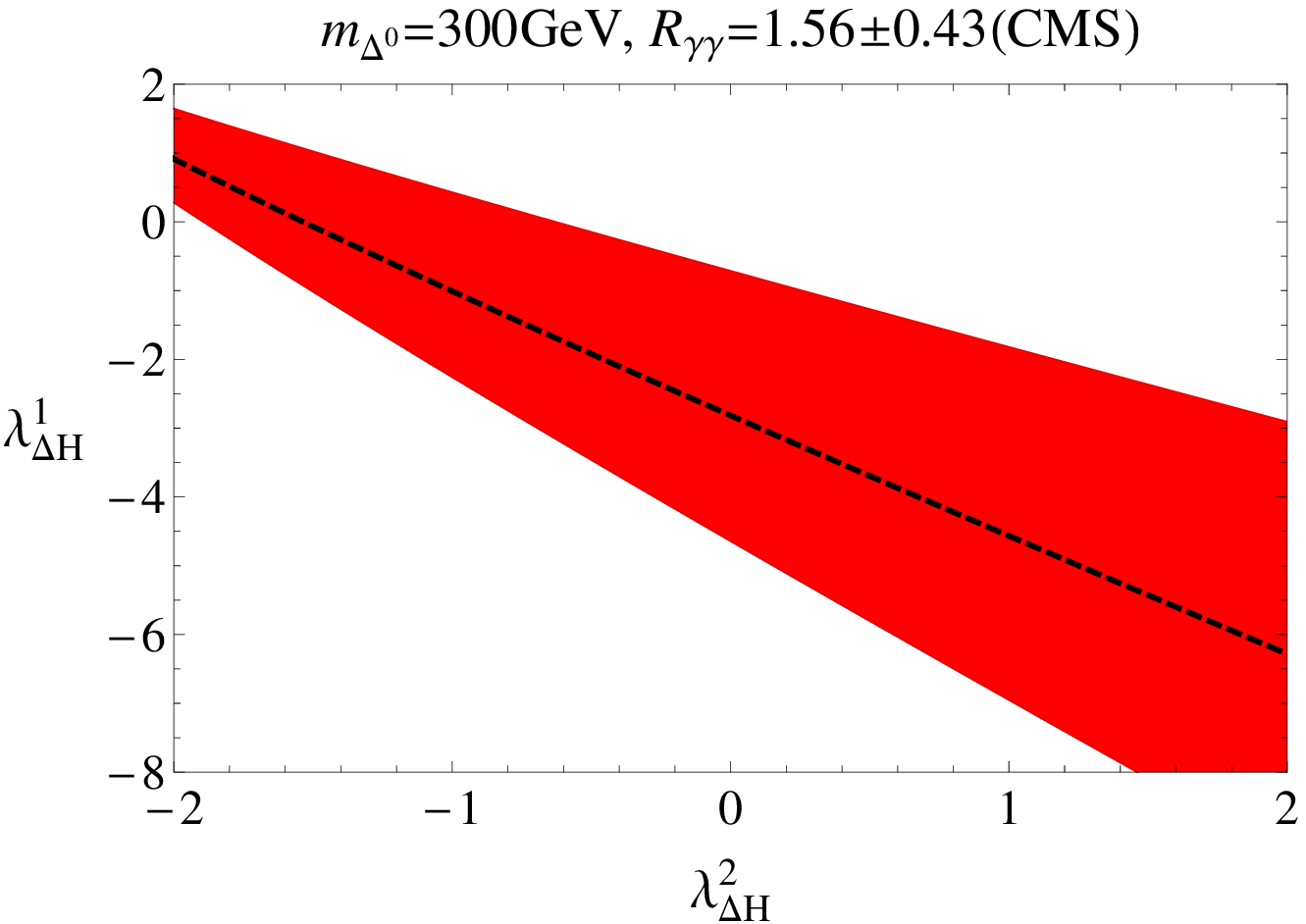}
\hspace*{2em}
\includegraphics[width=7cm]{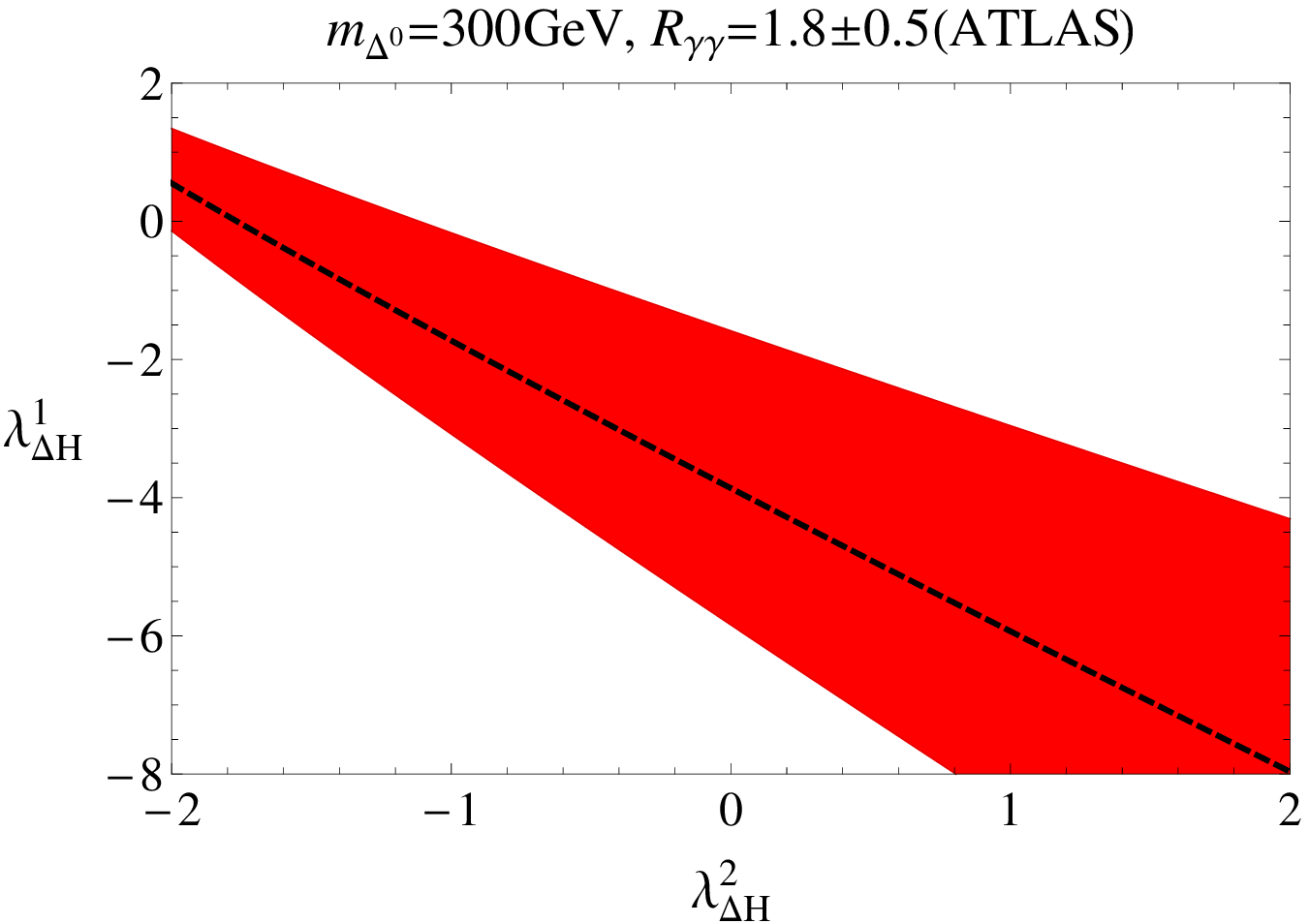}
\caption{Constraints on $\lambda^1_{H\Delta}$ and
$\lambda^2_{H\Delta}$ with $m_{\Delta^0}=300$GeV. } \label{h to rr}
\end{figure}

\begin{figure}[!h]
\centering
\includegraphics[width=7cm]{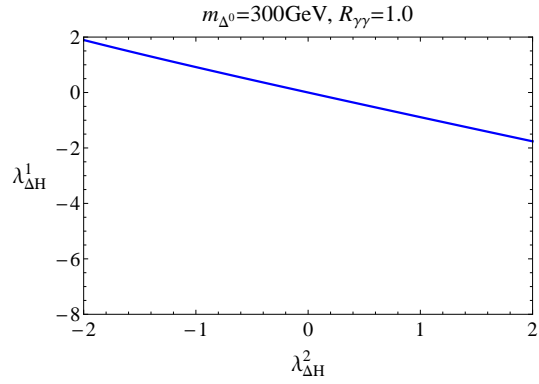}
\caption{Constraints on $\lambda^1_{H\Delta}$ and
$\lambda^2_{H\Delta}$ with $R_{\gamma\gamma} = 1$ for
$m_{\Delta^0}=300$GeV. } \label{h to rr-1}
\end{figure}

To see how the model can enhance the $h \to \gamma\gamma$ to be
consistent with LHC data, we will keep the $\Delta^{0}$ mass to
be $m_{\Delta^0} = 300$ GeV as used in the discussions on the neutrino masses and vary
$\lambda^{1,2}_{H \Delta}$  to obtain the new contributions to $R_{\gamma\gamma}$.
The results are shown in Fig.\ref{h to rr} and Fig.\ref{h to rr-1}.

 We also calculated new contributions to $h\to \gamma Z$. We
confirm the formalisms in Ref\cite{geng}. Using these formulas,  we
obtain the ratio of $h\to \gamma Z$ in the U(1)$_D$ model to that of
SM  as
\begin{eqnarray}
 R_{Z \gamma}
      &&= \vert 1- {2v \over A_{SM}^{Z\gamma}} \{{ g_{Z\Delta^{-} \Delta^{-}}(\lambda^1_{H\Delta} + {1\over 2} \lambda^2_{H\Delta})\over  m_{\Delta^-}^2} A_0(z_{\Delta^-},\lambda_{\Delta^{-}} )  \nonumber \\
      && + {2g_{Z\Delta^{--} \Delta^{--}}(\lambda^1_{H\Delta} + \lambda^2_{H\Delta}) \over  m_{\Delta^{--}}^2} A_0(z_{\Delta^{--}}, \lambda_{\Delta^{--}}) \}    \vert^2
 \label{R}
\end{eqnarray}
where $z_i = 4m^2_i/m^2_h$, $\lambda_i \equiv 4m_i^2/m_Z^2$ and $g_{Z\Delta\Delta} \equiv (T^3_\Delta - Q_\Delta s_W^2 )/s_W c_W$. $A^{Z\gamma}_{SM}$ comes from SM W boson and top quark contributions and $A_0$ comes from new charged scalars in this model. They are given by
\begin{eqnarray}
&&A_{SM}=\frac{2}{v}\biggl[\cot \theta_W A_{1}(z_W,\lambda_W) + N_c \frac{2Q_t (T^t_3-2Q_t s_W^2)}{s_Wc_W} A_{1/2} (z_t,\lambda_t) \biggr], \nonumber \\
&& A_{0} (x,y) = I_1(x,y), \nonumber \\
&& A_{1/2} (x,y) = I_1(x,y) - I_2(x,y),  \nonumber \\
&& A_{1} (x,y) = 4(3-\tan ^2\theta_W)I_2(x,y) + [(1+2x^{-1}) \tan^2\theta_W - (5+2x^{-1})] I_1(x,y),  \nonumber
\end{eqnarray}
where $T^t_3$ is the third component of isospin of top quark, and $I_1$, $I_2$ are given by
\begin{eqnarray}
&& I_1(x,y) = \frac{x y}{2(x-y)} +\frac{x^2y^2}{2(x-y)^2}[f(x^{-1}) - f(y^{-1})] + \frac{x^2y}{(x-y)^2} [g(x^{-1}) -g(y^{-1})],  \nonumber \\
&& I_2(x,y) = -\frac{x y}{2(x-y)} [f(x^{-1}) - f(y^{-1})] ,  \nonumber \\
&& g(x) = \sqrt{x^{-1}-1}\arcsin \sqrt{x}. \nonumber
\end{eqnarray}

\begin{figure}[!h]
\centering
\includegraphics[width=7cm]{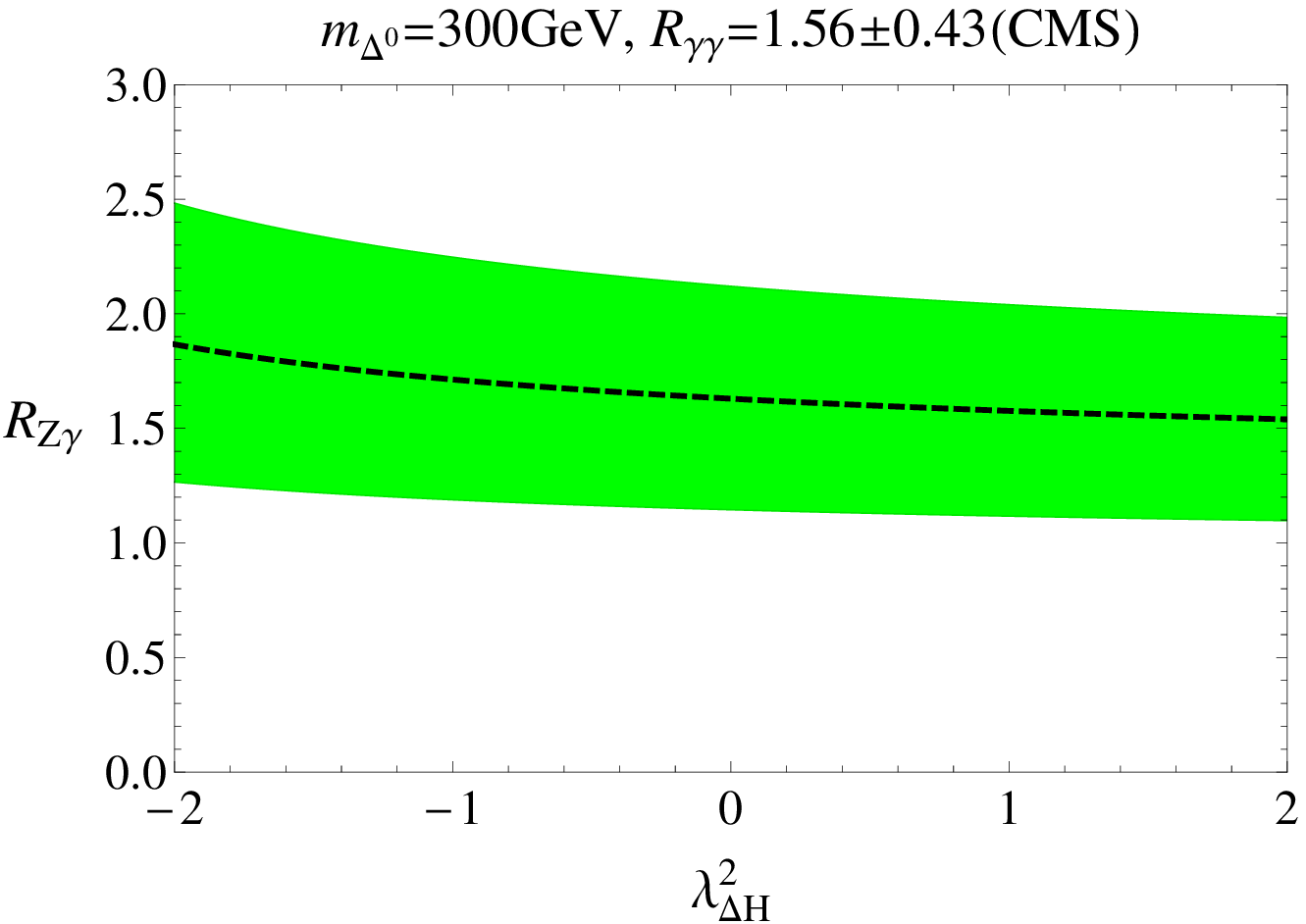}
\hspace*{2em}
\includegraphics[width=7cm]{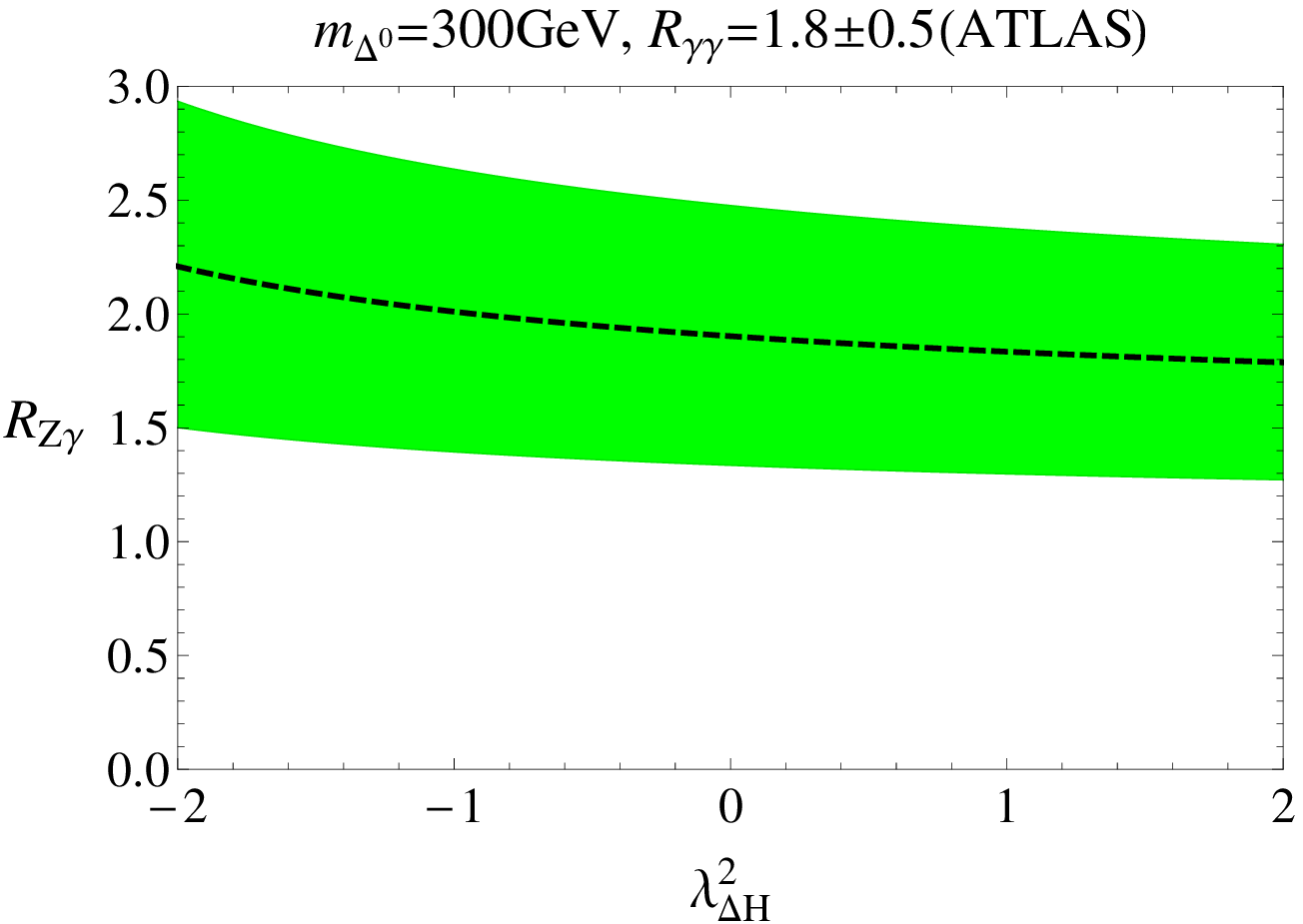}
\caption{Scaling factor for $h \to Z \gamma$ with the same
parameters for $h \to \gamma\gamma $.} \label{h to rZ}
\end{figure}

\begin{figure}[!h]
\centering
\includegraphics[width=7cm]{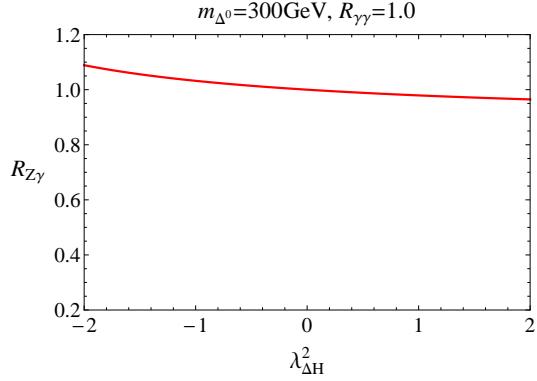}
\caption{Scaling factor for $h \to Z \gamma$ with the same
parameters for $R_{\gamma\gamma}=1 $.} \label{h to rZ-1}
\end{figure}

In the allowed
$\lambda^{i}_{H\Delta}$ space, $h\to \gamma Z$ will be modified significantly.
We
show the predicted scaling factor $R_{\gamma Z} = \Gamma(h\to \gamma Z)_{U(1)_D}/\Gamma(h \to \gamma Z)_{\rm SM}$ in Fig.\ref{h to rZ} and Fig.\ref{h to rZ-1}.

To enhance the ratio $R_{\gamma\gamma}$, negative
$\lambda^{1,2}_{H\Delta}$ are preferred. With fixed $m_{\Delta^0}$,
negative $\lambda^2_{H\Delta}$ implies that $m_{\Delta^0} >
m_{\Delta^-} > m_{\Delta^{--}}$. From Fig.\ref{h to rr}, we can see
that with negative $\lambda_{H\Delta}^{1,2}$ of order ${\cal O}(1)$,
the ATLAS and CMS results on $h \to \gamma \gamma$ can be
reproduced. If one controls the magnitude of
$\lambda^{1,2}_{H\Delta}$ as small as possible from perturbation
consideration, the optimal values for $\lambda^{1,2}_{H\Delta}$ are
around $-0.8$ and $-0.6$, respectively. With these values,
$\Delta^{-}$ and $\Delta^{--}$ masses are given by $284.5$ GeV and
$268$ GeV.  More negative $\lambda^2_{H\Delta}$ will make the mass
of $\Delta^{--}$ smaller which may be in conflict with LHC data. We
should take $|\lambda^2_{H\Delta}|$ as small as possible. With the
same parameters, the predicted value for $R_{\gamma Z}$ are shown in
Fig.\ref{h to rZ}.

With $\lambda^2_{H\Delta} > 0$, the mass hierarchy for the component
fields in the triplet is $m_{\Delta^0} < m_{\Delta^-} <
m_{\Delta^{--}}$. In this case, the new contributions may cancel out
if $\lambda^1_{H\Delta}$ is kept negative. We demonstrate this
possibility in Fig.\ref{h to rr-1} , where $R_{\gamma\gamma}$ is
kept to be 1. For this parameter space, the predicted $R_{\gamma Z}$
is shown in Fig.\ref{h to rZ-1}.

In general there is a correlation between $h \to \gamma\gamma$ and
$h\to \gamma Z$, that is, enhancement of $h \to \gamma \gamma$ leads
to an enhanced $h \to \gamma Z$.  This fact may be used as a test
for this model. Should an anti-correlation  between $h \to
\gamma\gamma$ and $h \to \gamma Z$ will be confirmed, this model
will be in trouble. But even if $h \to \gamma\gamma$ agrees with SM
prediction, $h \to \gamma Z$ can be different as can be seen in
Fig.\ref{h to rZ-1}.

Our analysis show that in order to explain the possible enhanced
$h \to \gamma\gamma$, negative $\lambda^{1,2}_{H\Delta}$ of order
minus one is needed. If the current data at the LHC will be further
confirmed, we need to check if the required negative
$\lambda^{1,2}_{H\Delta}$ are consistent with other constraints. One
of the constraints is from the stability of Higgs potential. Here we
argue that this is not a problem.

Potential bounded from below concerns potentials at fields taking large values. Let us consider terms involving
$\lambda_H$, $\lambda^{1,2}_\Delta$, and $\lambda^{1,2}_{H\Delta}$ in the case where $S$ and $\sigma$ fields are absent and carry out an similar analysis as in Ref.\cite{arhrib3}. At large values of $H$ and $\Delta$, the potential is given by
\begin{eqnarray}
\lambda_H x^2 + (\lambda^1_\Delta + \lambda^2_\Delta \eta) y^2 +  (\lambda^1_{H\Delta} + \lambda^2_{H\Delta}\xi) x y\;, \label{ff}
\end{eqnarray}
where $x = H^\dagger H$, $y = Tr(\Delta^\dagger\Delta)$, $\eta =Tr(\Delta^\dagger \Delta \Delta^\dagger\Delta)/(Tr(\Delta^\dagger\Delta))^2$, and $\xi = (H^\dagger \Delta\Delta^\dagger H)/(H^\dagger H Tr(\Delta^\dagger\Delta))$. The ranges
for $\eta$ and $\xi$ are $1/2 \sim 1$ and $0 \sim 1$, respectively.

 By definition both $x$ and $y$ are larger than zero. To satisfy Eq. (\ref{ff}), $\lambda_H$ and $(\lambda^1_\Delta + \lambda^2_\Delta \eta)$ must be positive. If   $(\lambda^1_{H\Delta} + \lambda^2_{H\Delta}\xi)$ is larger than zero, Eq. (\ref{ff}) is satisfied.

For $(\lambda^1_{H\Delta} + \lambda^2_{H\Delta}\xi) < 0 $ the
conditions are different.  This is the case if we use enhanced
$R_{\gamma\gamma}$ to explain the data, it is required that both
$\lambda^{1,2}_{H\Delta}$ to be negative are preferred. In the
following we study this case. The positivity conditions can be
obtained by requiring the diagonal elements and the determinant of
the following matrix to be positive
\begin{eqnarray}
M_p = \left (\begin{array}{cc}
\lambda_H&(\lambda^1_{H\Delta} +\lambda^2_{H\Delta}\xi)/2\\
(\lambda^1_{H\Delta} +\lambda^2_{H\Delta}\xi)/2&\lambda^1_\Delta +\lambda^2_\Delta\eta
\end{array}
\right )\;.
\end{eqnarray}

Without the conditions $\lambda^1_{H\Delta} <0$, $\lambda^2_{H\Delta} <0$,  $x>0$, and $y >0$, the parameters need to simultaneously satisfy
\begin{eqnarray}
&&\lambda_H >0,\;\;\lambda^1_\Delta + \lambda^2_\Delta >0, \;\;\lambda^1_\Delta + {1\over 2}\lambda_{\Delta}^2 >0\;,\nonumber\\
&&\lambda_H(\lambda^1_\Delta + \lambda^2_\Delta) >{1\over 4}(\lambda^1_{H\Delta})^2, \;\;
\lambda_H(\lambda^1_\Delta + \lambda^2_\Delta) >{1\over 4}(\lambda^1_{H\Delta} + \lambda^2_{H\Delta})^2\;,\\
&&\lambda_H(\lambda^1_\Delta + {1\over 2}\lambda^2_\Delta) >{1\over
4}(\lambda^1_{H\Delta})^2, \;\; \lambda_H(\lambda^1_\Delta + {1\over
2} \lambda^2_\Delta) >{1\over 4}(\lambda^1_{H\Delta} +
\lambda^2_{H\Delta})^2\;.\nonumber
\end{eqnarray}
In the above, we have also taken into consideration of the ranges of $\eta$ and $\xi$.

With the conditions $\lambda^1_{H\Delta} <0$, $\lambda^2_{H\Delta} <0$, $x>0$, and $y >0$, the conditions of positivity for Eq. (\ref{ff}) are relaxed to be
\begin{eqnarray}
&&\lambda_H >0,\;\;\lambda^1_\Delta + \lambda^2_\Delta >0, \;\;\lambda^1_\Delta + {1\over 2}\lambda_{\Delta}^2 >0\;,\nonumber\\
&&\sqrt{\lambda_H(\lambda^1_\Delta + \lambda^2_\Delta)} > - {1\over
2} \lambda^1_{H\Delta}, \;\;
\sqrt{\lambda_H(\lambda^1_\Delta + \lambda^2_\Delta)} >- {1\over 2} (\lambda^1_{H\Delta} + \lambda^2_{H\Delta})\;,\\
&&\sqrt{\lambda_H(\lambda^1_\Delta + {1\over 2}\lambda^2_\Delta)} > - {1\over
2}\lambda^1_{H\Delta}, \;\; \sqrt{\lambda_H(\lambda^1_\Delta + {1\over
2} \lambda^2_\Delta)} >- {1\over 2}(\lambda^1_{H\Delta} +
\lambda^2_{H\Delta})\;.\nonumber
\end{eqnarray}

A Higgs mass of 125 GeV, implies $\lambda_H = 0.13$. Our required $\lambda_{H\Delta}^{1,2} $ of order minus one  and the above conditions can be satisfied if one chooses both $\lambda^{1,2}_{\Delta}$ to be positive and satisfy $\lambda^1_\Delta + {1\over 2} \lambda^2_{\Delta} > 1/\lambda_H  = 7.7$ (with $\lambda^{1,2}_{H\Delta} = -1.0$). This condition can be easily satisfied by choosing $\lambda^{1,2}_{\Delta}$ to be about 5 which are well below the unitarity bounds on $\lambda^{1,2}_\Delta$ of order $4 \pi$\cite{arhrib3}.

Our model is more complicated because there are also $S$ and $\sigma$ fields. The term proportional to $\lambda_{\Delta \sigma H}$ can be chosen to be small and neglected. The corresponding $M_p$ matrix becomes a $4\times 4$ one. The conditions for potential bounded from below require the diagonal elements, the determinant of the matrix, and all determinants of its sub-matrices to be positive with the constraints for variables similar to $x$ and $y$ to be positive. The conditions for potential bounded from below include the ones discussed above, but have some additional ones. For our purpose, we need to fix $\lambda_H$ to be 0.13, and $\lambda^{1,2}_{H\Delta}$ to be around -1.0 to satisfy the positive conditions. Since several new independent parameters $\lambda_{S, \sigma, H\sigma, H S, \Delta S, \Delta \sigma, \Delta S}$ come into play, one is able to find reasonable parameter spaces to satisfy the conditions. For example, with $\lambda_{S, \sigma} >0$, $\lambda_{H\sigma,\Delta S, \Delta \sigma, \Delta S}$ to be zero. If one requires
$S$ to play the role of dark matter, $\lambda_{HS}$ should not be zero~\cite{two-loop inverse seesaw}. There is a large range for $\lambda_{HS}$ below 0.03 which can satisfy dark matter constraint for dark matter mass around half of Higgs mass and larger than 130 GeV~\cite{two-loop inverse seesaw}.  The positivity of potential at large values of fields can be satisfied.

 Before closing this section, we make some comments about
effects of the $\Delta$ particle at the LHC. In the case with
$\lambda^2_{H\Delta} < 0$, $\Delta^{--}$ is the lightest particle in
the $\Delta$ triplet. It is stable in the scenario where $2 m_S$
mass is larger than $\Delta$ mass, that is, $\Delta^{--} \to D^- D^-
\to l^- S l^- S$ is kinematically forbidden. This is the case for
the bench mark values we are using. With a mass of order a few
hundred GeV, $\Delta^{-,--}$ can be produced at the LHC with a cross
section of order about 10 fb. Although it does not decay into SM
particles making the direct detection difficult, being a stable
heavy charged particle it does leave tracks in the detector which
have been searched for at the LHC. The current data from LHC still
allow mass of order a few GeV\cite{charge_track}.  If it turns out
that $\Delta^{--}$ mass is large enough, and $\Delta^{--} \to D^-
D^- \to l^- S l^- S$ becomes kinematically possible, then $l^-l^- +
\slashed{E_T}  +$ jets is the signal to search. This has small SM
background and can be searched at the LHC.

In the case with $\lambda^2_{H\Delta} >0$, $\Delta^0$ is the
lightest particle in the $\Delta$ triplet. It can also be copiously
produced at the LHC because the mass can be as low as a few hundred
GeV. Search for this particle is similar to search for dark matter
which can annihilate into quarks. Some of the processes which can
provide information about this particle are single photon plus
missing energy and mono-jet plus missing energy.  ATLAS and CMS
experiments at the LHC have carried out such studies. At this moment
the data are not constraining enough to rule out the parameter space
we are using\cite{CMS_DM,ATLAS_DM,DM_constraints}. But as more data
become available, the model can be constrained more.

\section{ Conclusions}

We have studied some phenomenological consequences of a two loop
radiative inverse seesaw model with an unbroken
global $U(1)_D$ symmetry. This model has a natural candidate for dark matter which allows larger Yukawa couplings
and low mass of order a hundred GeV charged new particles in the
triplet scalar $\Delta$. The large
Yukawa couplings can lead to large leptonic flavor changing effects
in $\mu \to e \gamma$ and $\mu - e $ conversion. The current data
have already constrained the size of the allowed Yukawa couplings.
Future improved experiments on $\mu - e$ conversion can improve the
constraint by several orders of magnitude. The existence of low mass
charged particle in the triplet $\Delta$ make it possible to enhance
the $h \to \gamma \gamma$ to explain the deviation between the LHC
data and SM prediction.

\acknowledgments \vspace*{-1ex}
This work was supported in part by NSC of ROC, and NNSF(grant No:11175115) and Shanghai science and technology commission (grant No: 11DZ2260700) of PRC.

\end{document}